# Preliminary Study for Dosimetric Characteristics of 3D-printed Materials with Megavoltage Photons

**Seonghoon Jeong and Myonggeun Yoon**

*Department of Bio-Convergence Engineering, Korea University, Seoul, Korea 136-703*

**Weon Kuu Chung and Dong Wook Kim**

*Department of Radiation Oncology, Kyung Hee University Hospital at Gangdong,*

*Seoul, Korea 134-727*

In these days, 3D-printer is on the rise in various fields including radiation therapy. This preliminary study aimed to estimate the dose characteristics of the 3D-printer materials which could be used as the compensator or immobilizer in radiation treatment. The cubes which have 5cm length and different densities as 50%, 75% and 100% were printed by 3D-printer. A planning CT scans for cubes were performed using a CT simulator (Brilliance CT, Philips Medical System, Netherlands). Dose distributions behind the cube were calculated when 6MV photon beam passed through cube. The dose response for 3D-printed cube, air and water were measured by using EBT3 film and 2D array detector. When results of air case were normalized to 100, dose calculated by TPS and measured dose of 50% and 75% cube were 96~99. Measured and calculated doses of water and 100% cube were 82~84. HU values of 50%, 75% and 100% were -910, -860 and -10, respectively. From these results, 3D-printer in radiotherapy could be used for medical purpose accurately.

Dong Wook Kim and Myonggeun Yoon are equally contributed as corresponding author.

Email: joocheck@gmail.com, dwkim@khnmc.or.kr, wkchung@khnmc.or.kr

Fax: +82-2-440-7393

# I. INTRODUCTION

In general, when tumors are exposed to the high doses that are prescribed for a definitive or palliative goal, the surrounding normal tissues exposed to intermediate doses from primary radiation [1-4]. Thus, the goal of the treatment planning is optimized to identify the option that best satisfies two conflicting priorities: reducing the dose that the surrounding normal organ is exposed to, and focusing the prescription dose into a target volume [5-7]. To achieve this goal, high and complicate treatment technique were developed such as intensity-modulated radiation therapy, volumetric modulated arc therapy, tomotherapy, etc [8-11]. Also, some devices such as bolus or compensators have been used on the patient's skin to attenuate radiation beam for control of radiation dose in some cases. In addition variety immobilization tools for the patient setup to reduce the uncertainty from patient position and motion during the treatment [12-13].

Since, 3D printer were introduced in the industry as a main agent leading to the third industrial revolution, there are suggestions that some medical practices could be performed by making prosthetic appliances, recently. Researchers in Michigan University successfully installed 3D-printed structure which has exactly same shape with bronchus of Tracheobronchomalacia patient in left main bronchus of young patient [14]. For radiation therapy, there is report for the comparison between the traditional proton range compensator made by computerized milling machine (CMM) and 3D-printer based range compensator [15]. However, the dosimetric response of 3D-structures for megavoltage photons is very rare in spite of the 3D-structure has great potential for patient dose compensation or immobilization in the radiation treatment field.

In this preliminary study, we aimed to estimate the dose characteristics of the 3D-printer materials which could be used as the compensator or immobilizer in radiation treatment.

## II. EXPERIMENTS AND DISCUSSION

**3D-printing**

There are two production methods of 3D-printer: accumulating method and carving method. A carving method is carving materials to make something exactly same shapes with users want to make using laser. In this experiment, however, 3D-printer (Makerbot Replicator 5$^{th}$ generation, MakerBot Industries, Brooklyn, New York, USA) uses an accumulating method using extruder to melt PLA. An accumulating method is accumulating PLA from bottom to top. The printing resolution between each layer is 100, 11 and 2.5 micrometers for X, Y and Z direction. A density of structures is can be selected from 50% to 100% of original density. We made 3 cubes where each side has 5cm length using 3D-printer. Each cube was setted to build with different densities as 50%, 75% and 100% of original density. After finished the printing, measured masses of cubes were 33.6g, 25.5g and 146.8g, respectively. They were used to verify different dosimetric characteristics among different densities of products made by 3D-printer.

**CT scanning**

The images of CT images were obtained by CT (Brilliance CT Big Bore Oncology; Philips Medical System, The Netherlands. The HU values and their average in specific region were read and calculated by using ImageJ software (National Institutes of Health, Bethesda, Maryland, USA). Setup of each images are composed of experiment part and background part. As represented in figure 1, experiment part is filled with 3D-printed cubes, empty air for solid water equivalent phantom for each experiment. The volume of experiment part is 5cm x 5cm x 5cm (125 cubic centimeters). Other parts, named background part in this experiment, are composed of solid water equivalent phantom to consider scattered dose and Matrixx (iba dosimetry, Schwarzenbruck, Germany) to measure transit doses.

**HU values**

CT images of 50% case (left) and 100% case (right) are represented in figure 2. The experiment part is darker than its surroundings in the 50% case as there is nothing in the experiment part. In the 100% case, however, there is something similar to water inside the experiment part. The measured HU values of 50%, 75% and 100% were -913.7±36.4, -859.6±58.8 and -13.6±73.8.

**Data measurement and analysis**

The setup of data measurement is described in the figure 1. Experiments were set on the couch and beam was exposed from under the couch because solid phantoms beside the experiment part are crashed down if experiment setup would be up-side-down. The gantry of the 21iX Linear Accelerator (Varian Medical Systems, Inc. Palo Alto, Ca, USA) exposed from under the couch to stabilize the experiment setup. The 6MV photon beam with 10cm x 10cm field, 100cm SSD (source to surface distance) and 500 MU was exposed under the experiment part. 2D array detector (Martrixx, IBA dosimetry GmbH, Schwarzenbruck, German) could measure dose distribution which is passed through experiment part. We also used Gafchromic EBT3 film (International Specialty Products, Alps Road, wayne, USA) to measure dose distribution. EBT3 film was put between experiment part and Matrixx. EBT3 film was calibrated for photon beam which has 6MV energy, 10cm x 10cm field, 100cm SSD and 0~500 MU.

Dose distributions for each experiment were measured by Matrixx. We confirmed dose distribution and average dose passed through experiment part using OmniPro I'mRT software (iba dosimetry, Schwarzenbruck, Germany). We also obtained line profile of dose distributions to evaluate different characteristics between each experiment. MATLAB (MathWorks, Natick, Massachusetts, USA) was used to read dose information measured by the Gafchromic EBT3 film. The doses measured by Matrixx and Gafchromic EBT3 film were compared with the calculated results by treatment planning system (Varian Medical Systems, Palo Alto, CA, USA).

**Comparison of the measured results with calculated results**

The results measured by Matrixx were represented in figure 3 and figure 4. As shown as figure 3, a dose distribution of the air case, 50% case and 75% case shows similar dose distributions to each other. It shows the small square pattern inside the large square. Each small square represents the dose distributions of the radiation after passed through the experiment part. The measured doses in the center square were 1.15 ~ 1.21 times greater than the measured doses the side part of large square. The dose distributions of the water case and 100%, however, look different from the dose distributions of the cases which have small concentrated square in the center of the large square. They look large squares which have smooth plane without anything inside. The measured dose inside squares of the water case and 100% case were almost same with the dose in the side part of the air case, 50% case and 100% case. When the measured dose using Matrixx of air case was normalized to 100, the average doses of water case, 50% case, 75% case and 100% case are 82, 99, 96 and 84, respectively (See table 1). You can also see the line profile at the center of the field for each cases in figure 4. As represented in the dose distributions, peaks were detected in the air case, 75% case and 100% case. It is related with the small squares in those cases.

The dose distributions measured by Gafchromic EBT3 film were represented in the figure 5. The results of each case are very close to the results measured by Matrixx. The related average doses for each case were 100, 83, 98, 97 and 83. Treatment planning system also calculated the dose in the 2D array detector using CT images. The related average doses were 100, 98, 99 and 83 for the water case, 50% case, 75% case and 100%, respectively. You can see the comparison data in the table 1.

The dosimetric characteristics for the 3D-printing materials with 50%, 75% and 100% densities were verified for 6MV beam in this experiment. The dosimetric characteristics of 50% and 75% were very similar to the air and the dosimetric charcteristic of 100% cube is similar to the water. As represented in table 1, the results of 2 different measurements using Matrixx and Gafchromic EBT3 film were very similar to each other. It is also well matched with calculated results by treatment planning system.

There are small errors in the HU value results. The range of error is from 36.4 to 73.8. We concluded that the error in this experiment is negligible. Some researchers say there are small errors in HU values in the planning CT [16]. The measured average HU values in the planning have differences with reference value in some cases. But it would be negligible if the error is within 100. So the measured HU values for 3D-printing materials in this experiment could have confidence.

We could get the water equivalent products and the air equivalent products using 3D-printer. The dosimetric results in this experiment would be helpful to make the compensator or immobilizer in radiation therapy.

### III. CONCLUSION

The dose characteristics for the 3D-printing materials with various densities were verified for 6MV photon beam. The dose characteristics of 50% and 75% products were similar to the air while 100% product seems to be similar to the water. This information would give the guideline when we would make an immobilization tool which can play a role as compensator and real human phantom which can exactly describe inside the human body. This study was necessary for PLA based 3D-printer users who are planning to make something related to radiation therapy.

## ACKNOWLEDGEMENT

This work was supported by the General Researcher Program (NRF-2012R1A1A2003174), the Nuclear Safety Research Program (Grant No. 1305033) through the Korea Radiation Safety Foundation (KORSAFe) and Ministry of Food and Drug Safety (14172MFDS404), and Radiation Technology Development Program (2013M2A2A4027117), Republic of Korea.

Table 1. The dose comparison among the calculated dose by TPS and the measured doses by Matrixx and EBT3 film. The HU values of each case were also represented in the table.

|  | Matrixx | EBT3 | TPS | HU values |
|---|---|---|---|---|
| **Air** | 100 | 100 | 100 | -1000 |
| **Water** | 82 | 83 |  | 0 |
| **50%** | 99 | 98 | 98 | -910 |
| **75%** | 96 | 97 | 99 | -860 |
| **100%** | 84 | 83 | 83 | -10 |

Figure Captions.

Figure 1. The experiment setup and CT setup in this experiment. Experiment part was filled with 3D-printed cubes, empty air or water equivalent phantom. The solid water equivalent phantoms were put beside the experiment part to consider scattered doses. Matrixx was put on the experiment part to measure transit doses. Another solid water equivalent phantom were put on the detector to consider backscatter.

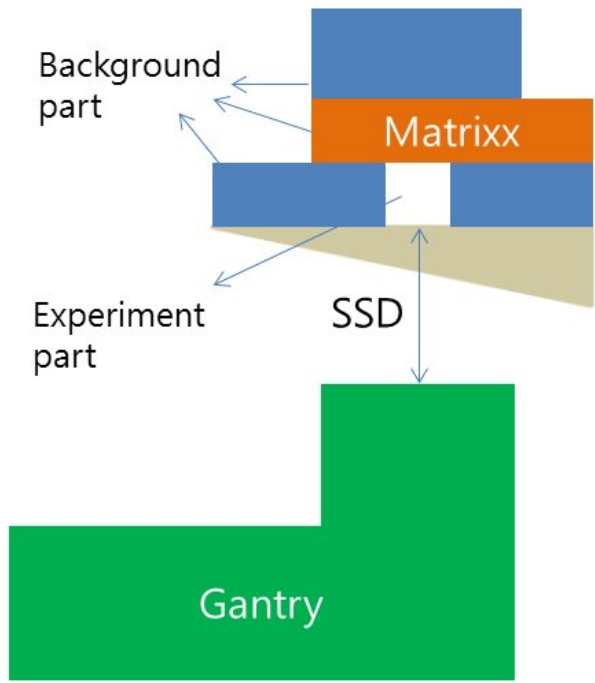

Figure 2. CT images of the experiment setup ((a) 50%, (b) 100%).

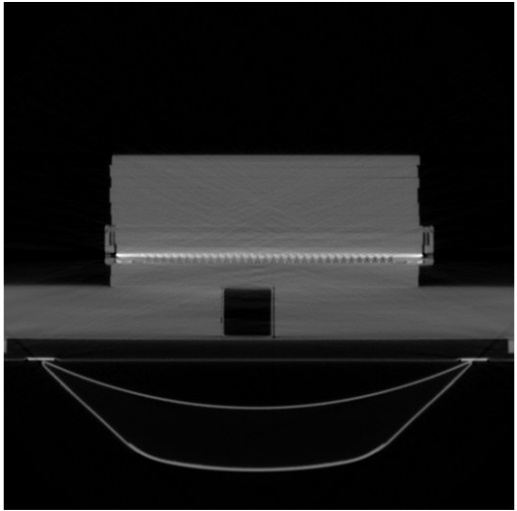 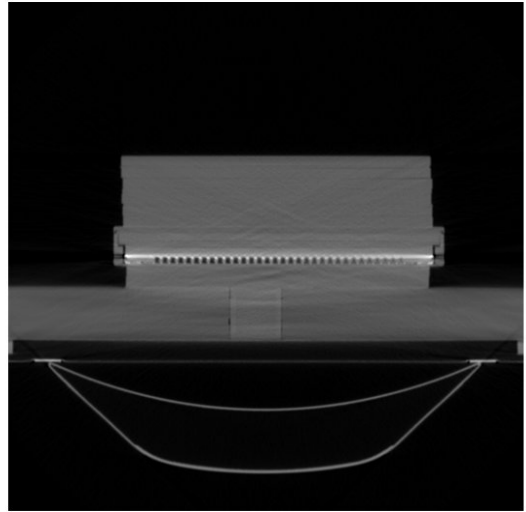

Figure 3. A results of dose distribution passed through the experiment parts ((a) air, (b) water, (c) 50%, (d) 75%, (e) 100%).

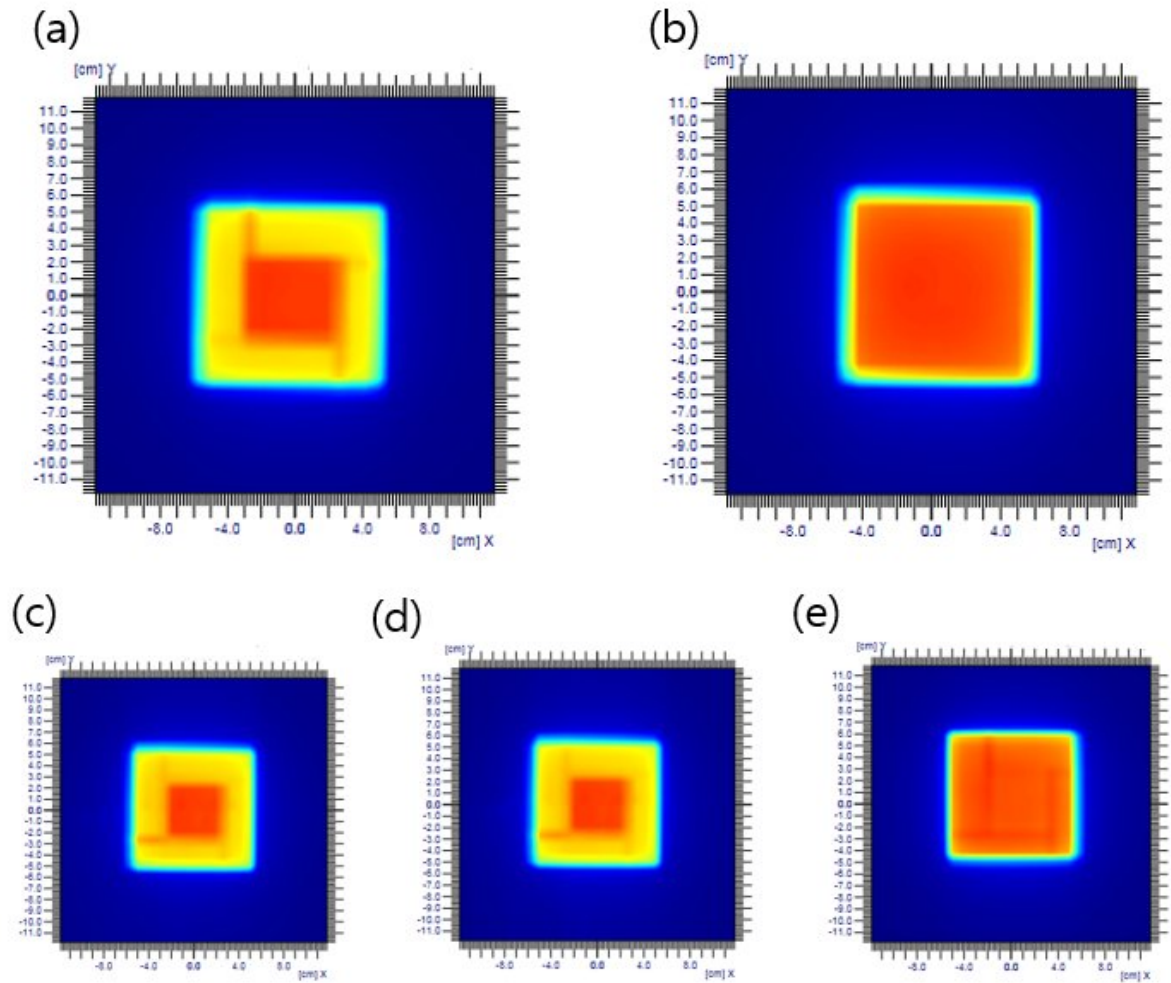

Figure 4. A results of line profile at the center of experiment ((a) water, (b) air, (c) 50%, (d) 75%, (e) 100%).

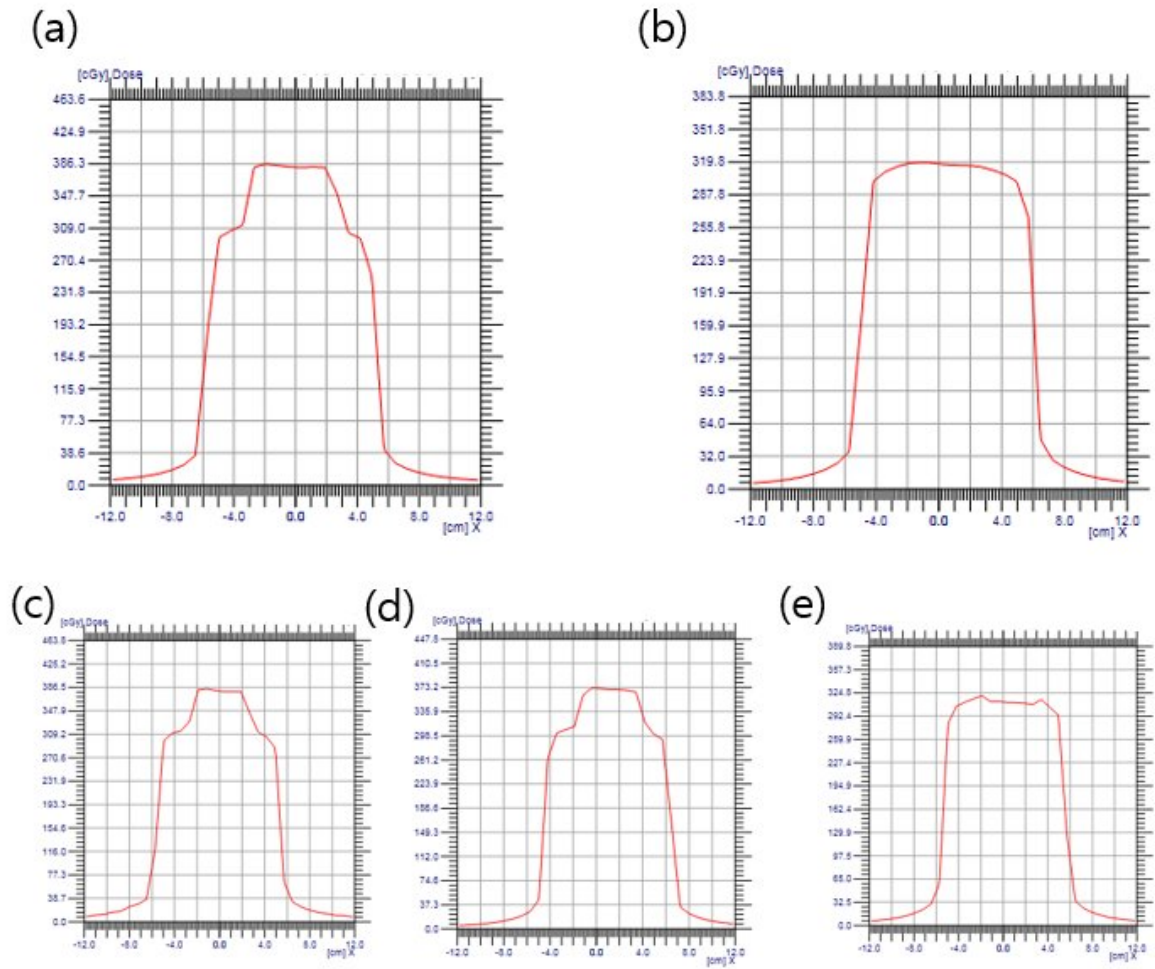

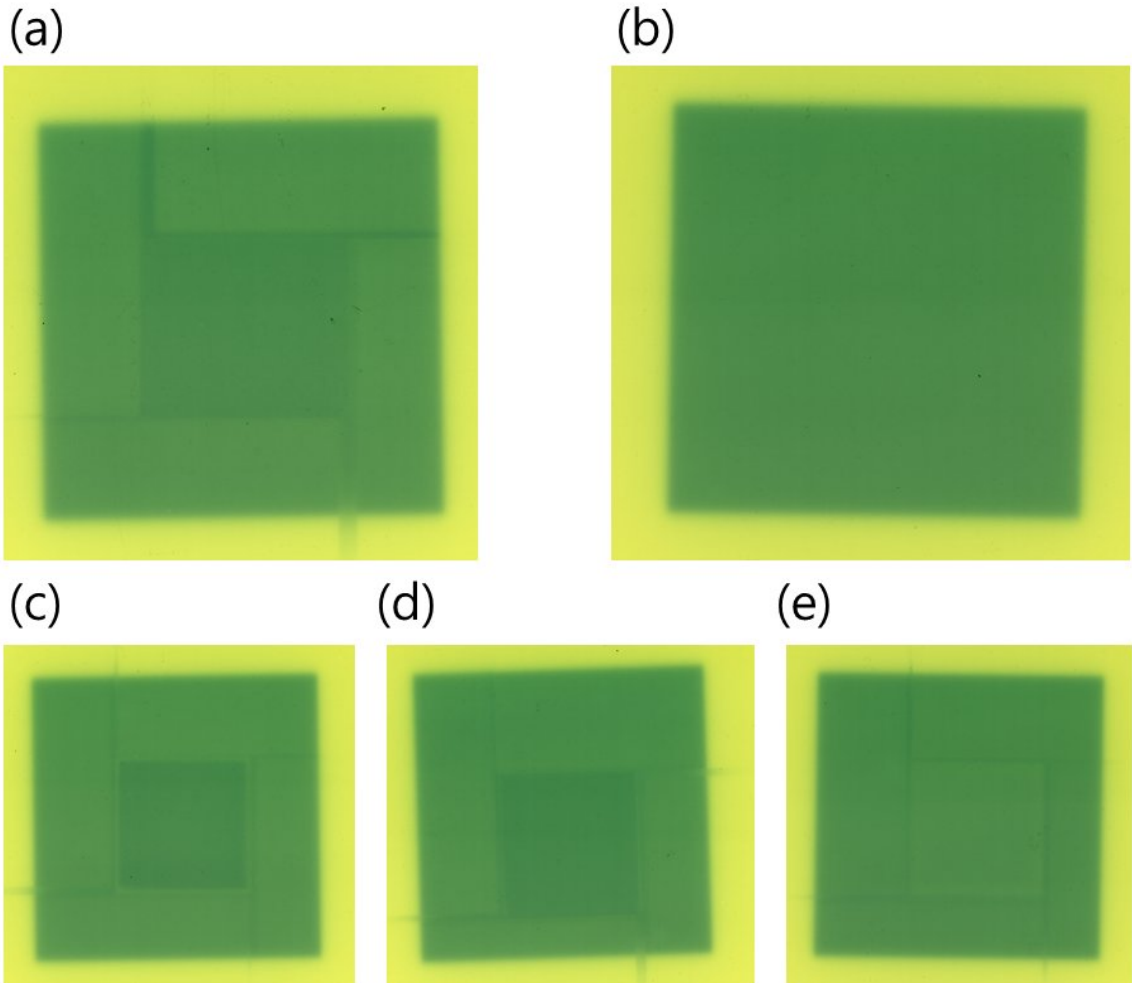

Figure 5. Dose distributions for each cases measured by the Gafchromic EBT3 film ((a) water, (b) air, (c) 50%, (d) 75%, (e) 100%).